\documentclass[]{hdsr}

\graphicspath{{figs/}}
\usepackage{booktabs}
\usepackage{makecell}
\usepackage{array}
\usepackage{paralist}

\begin{document}

% Larger bottom margin for the first page
\newgeometry{bottom=1.5in}

% Editorial staff will replace the following values:
% 1. Volume number
% 2. Issue number
% 3. Article DOI
% e.g. for Volume 2, Issue 3, DOI 12.345:
% \volumeheader{2}{3}{12.345}
\volumeheader{8}{3}{10.1162/99608f92.91b8834f}

\begin{center}

  \title{Unsafe at any AUC: Unlearned Lessons from Sociotechnical Disasters for Responsible AI}
  \maketitle

  % Start page numbering on second page. Must appear *after* \maketitle
  \thispagestyle{empty}
  
  \vspace*{.2in}

  % Authors and Affiliations
  \begin{tabular}{cc}
    Joshua A.\ Kroll\upstairs{1,*}, Andrew Smart\upstairs{2}, R. Stuart Geiger\upstairs{3}, Abigail Z.\ Jacobs\upstairs{4}
   \\[0.25ex]
   {\small \upstairs{1} Naval Postgraduate School} \\
   {\small \upstairs{2} Google Research} \\
   {\small \upstairs{3} University of California, San Diego} \\
   {\small \upstairs{4} University of Michigan, Ann Arbor} \\

  \end{tabular}
  
  % Replace with corresponding author email address
  \emails{
    \upstairs{*}jkroll@nps.edu
    }
  \vspace*{0.4in}

\begin{abstract}
  As automated decision-making and data-driven technologies pervade society and are used to manage consequential outcomes, understanding the technology's capabilities, limitations, and attendant risks in context requires analysis of full sociotechnical systems. Sociotechnical analysis of risks in highly complex systems provides clear lessons for the design and evaluation of AI systems, transcending a technical focus on reliable or ``responsibly designed'' components to understand risks at a systems level. 
  
  Human-made catastrophes have been studied for decades because of the severity of these events: consider Chernobyl, Three Mile Island, Fukushima-Daiichi, Bhopal, the Challenger disaster.
  A common misconception is that these kinds of events are freak
  accidents, resulting from the inherently unforeseeable interactions in complex systems. Closer examination reveals that the risks and hazards were well-known beforehand but not acted upon due to social structural, political and economic factors.

  We outline several areas where the development and use of AI can benefit from learning these unlearned lessons: improved risk perception, communication, and analysis at the organizational level; traceability of requirements and responsibilities; and holistic approaches to responsibility and safety that  include social and organizational dynamics as first-order engineering concerns. For each area, we offer concrete unlearned lessons and exemplify how they led to failure in prior accidents as well as examples of how these lessons remain unlearned for modern computing systems, particularly AI. 
\end{abstract}
\end{center}

\vspace*{0.15in}
\hspace{10pt}
  
\copyrightnotice

\section{Introduction}
In the 1986 televised hearings on the Challenger space shuttle explosion, Richard Feynman famously demonstrated before the U.S. Congress how a rubber O-ring seal would catastrophically fail in cold temperatures. Yet while the physical failure of this individual component was ostensibly the antecedent cause for the explosion, the reason why the Challenger exploded was much deeper: why was this risk missed? Such a failure mode and risk was known by some across the many different distributed teams at NASA and various contractors responsible for the shuttle. Diane Vaughan's landmark investigation identified many overlapping factors that contributed to a ``normalization of deviance'' --- a result of specific social, cultural, organizational, political, and economic conditions that obscured and undermined various safety concerns, including but not limited to the specific O-ring issue \citep{vaughan1996challenger}. 

Ralph Nader's 1965 \textit{Unsafe at any Speed} critiqued the auto industry-sponsored view that car crashes were idiosyncratic isolated incidents. Such incidents were explained away as human recklessness---the purview of individual drivers, traffic enforcement, and road design, not the automaker. While debates raged about speed limits, Nader reframed auto safety as a complex problem with interlocking responsibilities across society. (Seatbelts would help too.) Just as Nader argued there was no single cause and thus solution to auto safety, we too argue today's AI industry similarly can obsess over isolatable factors, technical fixes, or ritualistic practices. AI performance is often proxied by various metrics (like the Area Under the Curve or AUC that measures performance on benchmarks),
but high performance obscures the complex systems nature of safety.

In this most recent wave of AI, concerns around AI safety are rising --- often with the incorrect assumption that AI is a completely novel or alien technology, for which traditional concepts of safety will not work and must be reinvented from scratch~\citep[e.g.][]{yudkowsky_artificial_2008,bostrom_superintelligence_2014, hendrycks2021unsolved}. The multidisciplinary tradition of safety science has long analyzed such disasters across sectors, countries, and time, with one major common meta-level finding: safety must be approached as a complex systems problem, rather than focusing on individual technical components and technical solutions. No matter what the underlying technological components are, AI safety is just as social, organizational, cultural, economic, and operational as in classic complex socio-technical systems~\citep{raji2020concrete}. Safety science shows that risk governance must happen at the system-level, with specific interventions that can help mitigate risk --- but never `solve' it completely. 

We use \emph{``AI system''} to refer to a deployed, end-to-end socio-technical system, in part composed of learned models and automated routines that broaden the scope of tasks which can be automated usefully. Concretely, this includes technical subsystems (data collection and quality controls; data pipelines; feature generation; model training workflows; deployment infrastructure; user interfaces; logging and auditing; benchmarks and evaluation suites) as well as human and organizational processes (problem formulation and requirements analysis; metric selection and objective setting; review and approval processes; stakeholder engagement; procurement and system integration practices; individual and institutional incentives; cultural practices and beliefs; and political-economic forces). This definition is intentionally broad, as systems safety shows us how harms may befall humans from system behaviors as mediated through complex socio-technical systems.

Harms emerging from automated, data-driven sociotechnical systems are broad and extensively cataloged elsewhere \citep{shelby2023sociotechnical}, which occur for individuals and groups. Individuals can lose jobs, opportunities, basic freedoms \citep{propublica-bias}, and social services \citep{obermeyer2019dissecting}. Both individuals and groups can suffer representational \citep{noble2018algorithms}, dignitary \citep{wired-a-level}, and surveillance-based harms \citep{Leslie2020Tackling,diaz2024public}.

There is also much concern within AI development and across public media around so-called ``existential risk'' of AI, or apocalyptic scenarios around speculative future AI inventions and deployments, like Artificial General Intelligence. This concern is organized and influential, intersecting with the effective altruism and transhumanism movements \citep{ahmed2023building, gebru_tescreal_2024} in ways that ignore the lessons of safety science. Such an epistemic community has ``converged around a narrow set of technical solutions that do not engage with work that falls outside of the[ir] ideological and disciplinary boundaries'' \citep{ahmed2023building}. As such, we reject questions of risk ``from AI,'' as a disaster in itself, or from hypothetical future inventions --- especially when such framings draw attention away from how risk arises as components interact within specific, real-world complex systems.

We outline key unlearned lessons from these catastrophes about complex systems, political economies, risk communication, organizational culture, and safety. By examining the components of prior sociotechnical system failure, we gain visibility into organizational and social practices present in the development and deployment of AI systems which raise safety hazards. Reflecting on these lessons, we offer suggestions for how these lessons can inform the communities of AI research and practice. 
Our approach contrasts directly to existing broad thrusts of research on AI safety, which largely either aim to increase component reliability (better models, additional benchmarks, engineering safety features and metrics within AI models such as bias measures);  to enhance component performance requirements (alignment, fine-tuning of large models, robustness); or to undertake unvalidated risk-reduction rituals without consideration of how they lead to safe outcomes (model-focused auditing and documentation practices). These approaches treat technology problems as requiring solutions via more or better technology, adopting in the process a solutionist framing: assuming problems can be solved, rather than requiring continual effort and activity, management and repair. 

Instead, we turn to systems safety, which has long recognized the importance of social, organizational, and cultural practices and interventions to holistically help mitigate risks. We synthesize these lessons, which include cultural hazards like a permanent rush culture, 
misaligned incentives, symbolic compliance, or easily-gamed standards. We draw attention to more generalizable sociotechnical factors that can improve (but never solve) safety, including dedicated high-level roles for safety personnel, cultural acceptance and organizational incentives for curmudgeons (pathologically critical thinkers), critical self-evaluation after safety violations (or near misses), traceable processes, and more stakeholder-centered orientations and incentives.
Making AI safer cannot be solved with specific technical solutions or isolated practices:  safety depends on the entire complex system of people, organizations, incentives, and environments in which AI is designed, developed, and deployed. Drawing on lessons from a diverse range of fields including aviation, nuclear power, industrial manufacturing, and finance, we reframe the harms of AI as inherently produced by the interactions between humans, technological components, organizations, incentives, and governance structures at various scales.

\section{Systems Safety as an Analytical Lens}

The discipline of \emph{systems safety}~\citep{leveson2016engineering, leveson2020safety} arose as a response to increasing complexity in sociotechnical systems during the 20th century.
Safety engineering aims to identify ways a system might reach a set of defined, undesirable behaviors known as ``losses'' or harms, with loss events often referred to as ``accidents''. Systems approaches to safety view causes of accidents as emergent from system structure, transcending simple linear causal chains and scoping in as much of the development, operations, and management of a system as is necessary to assess its paths to defined losses. 

In contrast, engineering practice typically operates at the level of components and specifications for intended behavior, both in development and for analysis/assessment. Verification asks ``did we build the system right'', while validation asks ``did we build the right system?'' Both evaluate whether components behave according to spec, but neither generally captures how a complex system composed of those components will behave in real-world use. Specification correctness is defined by the observed behavior under assumed conditions, often modeled as simulated scenarios (e.g., fog for a self-driving car's vision models), and thus depends on anticipating the full scope of deployment scenarios. 
The study of accidents reveals inherently system-level problems, often retrospectively: in aviation safety, the phrase ``checklists are written in blood'' \citep{sull2015simple} captures that risk mitigation includes components (things to be checked) and organizational practices in the larger system (checklists)---but that it often takes disaster to implement and enforce such practices. 

It is tempting to believe accidents result from the failure to implement necessary precautions in a system or the failure for controls to engage --- whether because of technical failure or error by humans. However, this view oversimplifies safety: many accidents happen even when everything in a system works as intended. Humans involved in operating and managing a system should generally be assumed to be acting in good faith to avoid failure and to be making the best judgments they can with the information available to them at the time. When systems fail, it is because failure is a ``normal'', available behavior within the system; if designers had foreseen this behavior, the system would likely have been redesigned to eliminate its possibility~\citep{perrow1984normal}.

Systems safety recognizes that accidents arise from a combination of contributing causal factors, rather than a chain of explanatory precursors to failure. Such structural causation models can capture accidents that may not have a single root cause, tracing back cleanly to component failure or operator error~\citep{smart2024beyond}. In this way, safety control becomes part of an overall sociotechnical system. Hazards---system states which prefigure defined losses (i.e., accidents)---become visible targets for engineering methods, either redesigning the system to eliminate them entirely or designing controls ensuring the system will recover to a normal state~\citep{leveson2020safety}. A systems-level understanding of accident etiology enables recommendations for system-level interventions and improvements~\citep{delfos2024integral}.

Accidents can occur even when no components have individually failed~\citep{leveson2016engineering}. System behavior derives from the structure and hierarchy of components, their interactions, and the ways communication and control flows between components (including humans, often the most hazardous components)~\citep{checkland1981systems}. In AI systems, this distinction is routinely collapsed. Model-intrinsic properties like performance on benchmarks for ``fairness'', ``bias'', ``safety'' or ``alignment'', demonstrating robustness to input perturbation, or meeting high performance standards  are all component-level, extensionally visible---and thus testable---properties. These are necessary for responsible design, but they do not on their own establish safety. This is true even when functionality is concentrated in seemingly-monolithic components like large ``foundation models.'' Treating such a model as a single component may seem to reduce the interaction effects, but such architectures collapse a vast array of components, inputs, and human discretion into one component.

As we later detail, current approaches in AI Safety generally rely on technically-scoped approaches, which include: ``alignment'' or ``steering'' a model to produce desired rather than undesired behavior; benchmarks or practices like `red teaming' to probe for known failure modes; improvements to reliability or resilience to malicious input through adversarial training; new system components like filters or guardrails that check for unsafe behavior; or pushing risk perception and control down to the level of operators and users (post-deployment monitoring; interpretability)~\citep{bengio2025international, bengio2025international2}. Many of these approaches confuse safety and improving functionality, ultimately laundering responsibility for safety to front-line developers or to end-consumers of AI-enabled tools~\citep{dobbe2025ai}.

Safety science views safety as an \emph{emergent} property, meaning that it has a parsimonious description in terms of composed systems that is unnecessarily complex or difficult to express at smaller scales, such as in terms of components. Safety engineers attribute this emergence to the dependence and interactions between system components, including the fact that accidents can occur
even when components all follow their functional specifications. The concept of emergence comes from the theory of complex systems, where it refers to this notion of interaction-oriented effects and also to the idea that phenomenology at different scales may be described best in scale-dependent ways. 

Emergence helped safety science move beyond older conceptual models (such as the “Swiss cheese model,” in which failure is assumed to be random). As our analysis below shows, risk arises in ways that depend integrally on system structure and the key defenses against disaster must rely on this fact by focusing risk perception and reduction activities at the locations in the system where they are most meaningful. Better still is to design the system such that points of hazard do not exist, reducing the need for risk control altogether and creating “inherently safe” structures. Current research and development in AI under-attend this design-level approach to safety.

In the following section, we present a high-level taxonomy of organizational phenomena from the systems safety literature on disasters, pairing each with examples where research has shown it to be a contributing factor or precondition to sociotechnical disasters. 
Systems safety approaches generally distinguish between undesired outcomes (losses, accidents, and mishaps) and \emph{hazards}, system conditions which can lead to undesired outcomes but those outcomes have not yet come to pass. 

While not exhaustive, this taxonomy emphasizes the organizational and social practices present in the development and deployment of AI systems that present as hazards. We draw on primary and secondary literature from several salient examples of sociotechnical failures to derive a general high-level taxonomy of organizational errors and their recurrence that are particularly relevant for the AI field, referring to these categories of failure etiology as ``unlearned lessons''. We illustrate each unlearned lesson with several examples where research has shown it to be a contributing factor or precondition to failure.

\section{Unlearned Lessons}
\label{sec:unlearned}
The preconditions of disaster exist widely within the information technology sector. 
Workers and experts will recognize these factors: organizational cultures of ``no bad news''; permanent and relentless rush work; weak or deficient internal controls; short term financial thinking; poor risk perception or active concealment of risk; poor communication; and a belief in technical solutions to human concerns (or `solutionism' ~\citep{vinsel2020innovation}).

Our analysis maps known challenges of systems safety to the assessment of real world AI systems. It is tempting to break AI systems into a general architecture: e.g., data collection \& cleaning; model training; model evaluation and qualification; model deployment, serving, and monitoring; logging and auditing of queries; feedback to new versions. One could similarly build a process/lifecycle model based on activities undertaken by developers. Our analysis depends on neither, however, so we leave such a model unspecified in order to avoid dependence on unnecessary details. From systems safety, we can theorize about risk and its management at the system level, regardless of the specific component breakdown of a given system-under-evaluation. For real world AI, this reveals that AI systems are not so novel nor resistant to assessment.
To deploy AI technologies responsibly---to prevent both everyday misbehavior and large-scale crises---the AI research and development community must follow other high-risk domains in appreciating the crucial nature of managing social, organizational, and cultural processes (in addition to technical measures) when defining and assessing safety. 

\subsection{Poor Risk Perception}

Engineering and maintaining large systems requires estimating risks throughout the design and operations processes. Often, organizations systematically underestimate risk. The safety literature suggests a variety of causes: failing to account for interacting component risks; risk concealment or fragmentation throughout the design process (whether due to hubris or in response to incentives); or limited imagination around what could go wrong \citep{leveson2016engineering, leveson2023introduction}. For the latter, the assumptions about possible system behaviors can easily describe a set of behaviors smaller than those reachable by the real system \citep{leveson2004new}.

Even when identified, failures and risks are often calculated as the product of a series of probabilities of individual component failures, assumed (or asserted) to occur with low probability, and ignored as a result~\citep{perrow1984normal}. Taking an assertion of  low-probability failure as given (especially from the component’s manufacturer) gives a false sense of control. 
Such an approach also tends to rely upon technical assertions of component failure probabilities when properly maintained. Maintenance is not well valued or factored into risk models. This is true both in the sense of continuance operations---keeping machines in a functional state---and of maintaining systemic discipline in operations---following procedures even when they seem unnecessary~\citep{russell2018after}. The result is often shorthanded as the ``drift towards failure''~\citep{dekker2016drift}. 

The resulting illusion of control---paired with a fragmentary perception of risks {across an organization, across a system}---leaves stakeholders susceptible to catastrophe: The Challenger explosion remains a canonical example of such failure to integrate knowledge of risks~\citep{vaughan1996challenger}. 
At a different scale, the 2008 financial crisis followed a similar pattern, where a period of irrational exuberance was paired with new financial technology that effectively created an incentive and mechanism to under-rate and hide risk~\citep{nelson2014uncertainty, arora2011computational}. Meanwhile, the case of the Therac-25, a hospital radiation machine that led to multiple deaths and injuries in the 1980s, revealed how overconfidence in design safety and poor risk assessment by operators and regulators caused incidents to be systematically ignored or discounted~\citep{leveson1993investigation}.

In AI systems, one widely-cited example \citep{obermeyer2019dissecting} was a healthcare machine learning model used to allocate care for high-risk patients, which used actual healthcare costs as a proxy for each patient's medical severity. Because structural inequalities in the US mean Black patients spend far less on healthcare for equivalent conditions, the model systematically underestimated Black patients' health needs. While the developers were explicitly concerned with the risk of bias, they failed to perceive that their choice proxy variable obscured this risk.

\subsection{Incentives}
The motivations of people -- whether they are setting requirements, designing, operating, or making decisions within a system -- often create the conditions for failures. Organizational cultures shape how these motivations drive actions, whether it is taking additional risk to achieve a performance metric or to drive growth or extend a business relationship. Hazardous incentive-driven behavior includes performing the actions associated with risk management as an end in themselves, as well as decoupling valid risk perception or control activities from their actual risk reduction functions~\citep{power1999audit}.  Incentive effects drive organizational culture in ways that create a loss of accountability, especially by refocusing effort toward functionality improvement and away from safety or risk assessment, activities often seen to add cost, delay projects, and limit the speed of functionality improvement.

Business pressure caused Boeing, for example, to suppress many safety activities surrounding the addition of a new subsystem in its 737 MAX-8 aircraft. They argued to regulators that the successful performance of that subsystem in an existing military aircraft, the Boeing KC-46 air tanker, was sufficient evidence. However, in the context of the MAX-8, the system had key differences that were not communicated to pilots, which led to a series of fatal accidents and the grounding of the entire fleet for well over a year 
\citep{herkert2020boeing}. 

\subsection{Permanent Rush Cultures}
Especially in high-tech work cultures, there are heavy pressures to complete work on schedule and under budget, so when problems cause delays, the shortfall often is salvaged from safety margins or other necessary care. This “rush work” stance is often cited as a cause for accidents. For example, since the US Navy introduced its `SUBSAFE' program to assure quality in hull integrity work, the only covered ship which has been lost in an accident was the \emph{USS Scorpion}, after its planned maintenance overhaul was reduced in scope and time by eliminating safety-critical work to put it back in service faster due to Cold War operational pressures~\citep{leveson2012subsafe}.
\looseness=-1

A similar rush culture has been identified in the former Soviet Union as contributing to the Chernobyl disaster. The Politburo demanded to increase the speed of construction of nuclear power plants to meet urgent domestic energy demands, and this constant haste lead to the nuclear industry in the USSR ignoring hazards, safety shortcomings, and design flaws~\citep{chernov2016man}. Risks to reactors were systematically downplayed as insignificant. Proximate to the accident, a planned safety test was delayed many hours into the night so that power output could remain high to support the meeting of end-of-quarter production quotas at nearby factories, causing a hazardous condition unrecognized by the late-shift operators.

As in other domains, computational systems including AI are being rapidly developed and deployed~\citep{sculley2018winner}. While speed-to-market is a key pressure in contemporary tech capitalism~\citep{styhre2020political}, classic software engineering management famously discovered that adding more programmers to a project often slowed down development, as the complexity of coordination increased ~\citep{brooks1975mythical}. Organizations that develop AI systems tend instead to prize speed and technical novelty in fear of missing out to their competitors, even going as far as to explicitly eschew the already-minimal discipline within a ‘move fast and break things’ tech culture.

\subsection{No Bad News}
Organizational culture can lead individuals to hide ``bad news'', especially when it might affect output or if no solution is available. This leads employees to keep secret information which might have averted disaster. The risk department of Silicon Valley Bank performed an internal stress test using its own risk models, which revealed vulnerability of the bank's bond portfolio to a rise in interest rates. This finding led the risk organization to change core assumptions in the bank's risk model relating deposit income to changes in interest rates. Approximately 2.5 years later, sales of bonds at a loss tied to rising interest rates were the proximate inciting factor to a run on the bank's assets that led to its collapse~\citep{waposvb}. Incentives for short-term profit, coupled with a culture that rejected information which threatened desired business growth, led the bank to accept more risk than it could bear.

Big tech is no different. Infamously, Meta's chief operating officer named her personal conference room at the company ``Only Good News''~\citep{wolverton2018onlygoodnews}, projecting a culture in which problems with the company's products were subordinate to revenue and growth. The consequences are well-documented and multifaceted. Former Meta employee Frances Haugen testified that when internal researchers found Instagram's engagement algorithms had significant mental health impacts on teenage girls in particular, findings were systematically ignored \citep{haugen_protecting_2021}. 

\subsection{Safe Components Do Not Imply Safe Systems}
Disaster often results even when everything is working `as expected'. System failures can arise from aggregations of components operating correctly due to unforeseen interactions~\citep{perrow1984normal}.
In the Chernobyl disaster, operators insisted to investigators that they had followed established safety procedures faithfully. The disaster resulted not from any single component failure, but from an engineering culture that enabled engineers to operate the reactor in ways designers did not intend, making seemingly insignificant changes to components of the reactor's control rods with inadequate records~\citep{plokhy2018chernobyl}. Investigation did not uncover specific component failures or operator deviation from procedure, even after the reactor's state had departed the normal operating envelope---instead, the accident depended on interactions between components operating as designers had imagined. Such unintended interactions can be exacerbated by component failure or by unforeseen component behavior, so component verification \& validation is a necessary but insufficient precondition for system assessment.

The assumption that machine learning is an entirely novel and unique form of software engineering has resulted in widespread abandonment of traditional requirements engineering for software~\citep{hutchinson2021towards,sculley2014machine}. However, seemingly magical (mis)behavior in traditional hardware and software systems is not a new, AI-specific phenomenon. One technique for supporting safety arguments is the formal verification of an implemented system against its engineering specification - for software systems (including AI-based ones), this can close the gap between assessment of the real artifact and of the specification in abstract. However, all such verification requires strong assumptions. When systems are complex and poorly understood, unexpected behavior is just a specific kind of surprise, not evidence of inherently new capabilities. Beyond over-strong assumptions, verification that a component meets its specification does not speak to the implications of interaction between components, even when they meet those assumptions.

For hardware components, transmogrification - the spontaneous transformation of one kind of part to another - has long been identified as a source of behavior that results from relatively simple device physics, such as bit flips from cosmic rays. In Space Shuttle mission STS-124, a crack in a sensor circuit converted a diode into a capacitor and caused each control module to interpret the sensors differently~\citep{driscoll2003byzantine}. 

Many consequential failures can be identified via building sufficient test \& evaluation plans to cover known paths to failure (e.g., the Intel “f00f” and “FDIV” bugs, which complete testing would have revealed). In other cases, surfacing assumptions about how components operate is a necessary antecedent for safety assessment (e.g., cybersecurity ``speculative execution'' leaks were a foreseeable consequence of data owned by parallel-operating executing code). In either situation, even full knowledge that a component operates correctly is not sufficient to build claims about the safety of the full composed system.
The history of complex traditional software systems (operating systems, databases, concurrent systems, distributed computing systems) is full of anecdotes in which surprising system behavior undermined design goals or created situations where component behavior was insufficient to describe system-level behavior. For example, improperly formatted updates to security scanning data in a popular security scanning software tool caused a large fraction of the world's computers to fail-stop and require rebooting, causing general breakdown in transportation networks, financial markets, and other critical infrastructure~\citep{nolan2024consequences}. 

\subsection{Kick the Can}

Humans in the system (e.g., operators) often serve as the locus of blame when technical components functioned as intended, whether or not they had the capacity to prevent failures~\citep{elish2015praise, elish2019moral}.
Too often, system design pushes difficult decision-making downward onto operators (or users of an AI system) as a way to restructure responsibility for failures off of designers and controllers so as to claim that bad outcomes were the result of misuse or failure to follow procedures. This echoes the early history of safety analysis as workplace safety, where mechanical automation in factories caused many injuries. Both factory owners and machine designers blamed these injuries on worker misuse. The eventual adoption of design principles for reducing risk, such as ``by designing the safeguards on machines and equipment so that if [an operator’s] acts are essential to safety, it becomes mechanically necessary for [them] to perform this act before proceeding with his task''~\citep{hansen1915standardization}, reflect the truth that responsibility for accidents vests with the system as a whole, requiring appropriate interplay between design and operator actions.

The problem of determining responsibility when harms to humans occur from algorithmic systems requires moving beyond simply blaming the poor decisions of human operators, who may not have meaningful control over the system nor have all the relevant information necessary to prevent accidents. For example, in AI development responsibility gets laundered off even the AI developers by introducing the problem of many hands~\citep{cooper2022accountability,kroll2020accountability}, whether through system complexity (which makes assignment of responsibility unclear, at conflict, or simply difficult) or because failures are treated as inevitable and unpredictable by the organizational culture. 

For example, a famous mid-air collision accident concerns the rules about whether pilots must follow guidance from an automated system recommending actions when an aircraft detects nearby traffic, or whether these recommendations can be overridden by pilots or controllers.
This gap can complicate the operationalization of control during technology procurement, as the lines of responsibility for aspects of avoiding harm are unclear~\citep{mulligan2019procurement, kroll2022responsible}.

Organizations may also attempt to shed responsibility by pushing decisions about how a technology integrates into even high-risk activities down to the discretion of involved personnel (human in the loop), replacing structural analysis of the suitability of technological interventions with the professional judgment of workers~\citep{elish2020repairing}. This move shifts responsibility: a medical chatbot that recommends contraindicating medicines is not blamed; the nurse whose job has become the human in the loop is instead blamed for not catching it.   Workers may be experts (medical or legal professionals, military commanders) and thus afforded substantial discretion and control. But because technology reifies a set of exogenously imposed policies and decision choices onto professional judgments, workers are forced into a situation where they must ``repair'' technology into their workflow---closing gaps between the rule-driven approaches to work embedded in the technology and their professional outlook. Filling the space between professional expert judgment and the rigidity of tool behavior requires professionals to make use of information on how to manage risk effectively, information often not provided or even obscured by the tools themselves.

We see echoes of this across domains.  In 2018 an Uber automated driving system failed to recognize a bicyclist, whom it struck and killed. The safety driver in the car was charged with negligent homicide, even though the automated system also failed to identify the bicyclist. The human driver here functions as a ``moral crumple zone'' that absorbs responsibility for a bad outcome when in fact the human operator is part of a larger complex system that failed, whether or not the operator’s agency contributes to avoiding failure~\citep{elish2019moral}. Employing humans as backups to automation is generally ineffective because not being in active control leads to complacency and degradation of situation awareness \citep{bainbridge1983ironies}. This was clearly the case in the Uber accident where in-car footage revealed gaps in the safety driver's capacity to control the vehicle as a backstop. We see the same moral crumple zones appear in explanations of Chernobyl, where the plant's human operators absorb much of the blame for the accident~\citep{wapochernobyl, plokhy2018chernobyl}. 

Similarly, the pilots of Air France flight 447 were blamed as the primary cause of the crash. Elish argues: ``the primary causes of the accident are found in the interactions \textit{between} automation and human, there are no certifications that cover this. Because the autopilot did not malfunction in a way recognized through its certification process, the only possible malfunction, systemically, is the human pilot, becoming the moral crumple zone''~\citep{elish2019moral}. Dekker posits that safety is not improved by assigning functions to humans or to automation based on which is better, but rather based on helping the two approaches to control ``get along''~\citep{dekker2002maba}.

\begin{table}[H]
\caption{Unlearned Lessons from Sociotechnical Disasters}
\label{tab:unlearned}
\renewcommand{\arraystretch}{1.2}
\begin{tabular}{|p{2.25cm}|p{4.5cm}|p{3cm}|p{4.5cm}|}
\hline
\textbf{Unlearned Lesson} & \textbf{Symptom} & \textbf{From Disasters} & \textbf{Expression in AI Systems} \\ \hline

\textbf{Poor \newline Risk \newline Perception} 
& Organizations underestimate risk through bad assumptions, ignoring interactions, information fragmentation, information concealment, or normalizing poor system hygiene and a drift towards failure.  
& Challenger (loss of understanding of safe temperature envelope); \newline Three Mile Island (fragmentation of maintenance state knowledge) 
& Benchmarks (bias, toxicity, alignment) treated as safety; risk framed as a model property, not use-context \\ \hline

\textbf{Incentives} 
& Growth, profit, scale, and schedule pressures override other concerns; symbolic risk management with little accountability; no whistleblower protections 
& Boeing 737 MAX 
& org structure / OKRs reward releases, growth, and scaling; safety review lacks veto power; shipping beats mitigation \\ \hline

\textbf{Permanent Rush \newline Cultures} 
& Safety margins sacrificed for speed; hazards ignored under pressure; organizational discipline eroded by urgency 
& Chernobyl; USS Scorpion 
& ``AI race'' framing; rapid model/product cycles; betas shipped as infrastructure; minimal stabilization \\ \hline

\textbf{No \newline Bad \newline News} 
& Risk information suppressed; discouragement of dissent; organizational hostility to problem identification without immediate solutions 
& Silicon Valley Bank collapse; Facebook 
& Harms research sidelined; findings routed through legal/PR; risk reframed as reputational \\ \hline

\textbf{Safe Components $\not\Rightarrow$ \newline Safe \newline Systems} 
& Failures emerge from interactions among correctly functioning components; safety cannot be inferred from component reliability or specification conformance 
& Chernobyl; Many cybersecurity incidents
& Safety delegated to guard models/filters/prompting; model certification substitutes for system governance; benchmarks treated as safety \\ \hline

\textbf{Kick \newline the \newline Can} 
& Responsibility displaced onto operators and users; accountability diffused through system complexity; humans function as moral crumple zones 
& Uber autonomous vehicle crash; Air France 447; Chernobyl  
& ``Human-in-the-loop'' shifts liability to users/contractors; providers disclaim downstream harms \\ \hline

\end{tabular}
\end{table}

\section{Mitigating Risk Sociotechnically, not Instrumentally}

Drawing on the systems safety science literature and the unlearned lessons identified above, we present several sociotechnical factors that contribute to AI system safety. We do not dismiss the large body of work on technical improvements in AI tool performance and reliability. Instead, we contextualize the value of such technically-focused work within a system-level approach to risk perception and management, to understand when and how it contributes to safer AI applications in real-world use.

We organize this section around an intentionally broad, sociotechnical understanding of \emph{safety cultures} \citep{silbey2009taming,guldenmund2000nature}. Organizations may attempt to counter hazards like those in Section~\ref{sec:unlearned} by cultivating safety-oriented norms through training, policies, structures, incentives, and sustained participation in well-tested organizational processes. We begin with the caution that governs the rest of this section --- that a safety culture is not an instrument to be deployed  --- and then survey the practices and qualities commonly associated with safety cultures. We then examine recurring themes from the literature, including: curmudgeons and critics; traceable processes; inclusive and psychologically safe spaces; diverse teams; fostering meaningful external participation; and treating failure as normal and expected.

\subsection{Culture is not an operating system}

Before turning to specific practices, we warn against the instrumental trap of treating a safety culture as a component-level `solution'. Safety cultures emerge from organizational and social dynamics, and their elements are not deployable artifacts with predictable risk-reduction effects. Silbey warns against this, noting that engineers often treat `culture' as though it were an instrumental technology, like an operating system to be installed~\citep{silbey2009taming}. Culture is instead a historically produced, power-laden field of norms, meanings, and practices through which people make sense of sociotechnical systems and grasp for the `right' ways to act within them. Culture conditions which safety actions are available, intelligible, and legitimate, not a mechanism that performs those actions itself. Some team-training practices can substantially improve safety when they enable operators to overcome barriers such as hierarchy and rigidly assigned roles, as well as to develop shared mental models of system state and task performance~\citep{salas2008does, mathieu2000influence}. Culture is a permitting frame, one that allows a junior operator or pilot to override a senior one when safety demands it, especially when incentives are not aligned. However, whether the override holds depends on the organization having decided to take a short-term loss to avoid a larger catastrophe. The dynamics identified in Section~\ref{sec:unlearned} help produce and reproduce this cultural context, often eroding the conditions that make safety action possible.

\subsection{Common safety culture practices}

Research on ``high-reliability organizations'' shows that strong safety performance can sometimes be sustained despite known limitations or unreliability in technical components, when those limitations are actively managed through organizational practices~\citep{roberts1990managing}. Such organizations tend to use checks and balances to detect and mitigate unexpected conditions, value maintenance and sustainment work rather than disproportionately rewarding technical novelty or ``innovation''~\citep{vinsel2020innovation}, and treat anomalies, degraded states, and small irregularities as conditions requiring attention rather than as tolerable deviations from normal operation. These organizations also distribute safety responsibilities broadly while assigning decision-making accountability clearly. They investigate failures in order to improve the system rather than treating individual blame as the primary remedy, an orientation commonly described as a ``no blame'' or ``blameless'' culture~\citep{wachter2009balancing}. They may also introduce redundancy and partially overlapping responsibilities to enable independent cross-checking, while attending to the new coordination and interaction risks that redundancy may introduce. 

High-reliability approaches move beyond purely technocratic visions of safety by making operational, organizational, and experiential knowledge relevant to risk perception and control, rather than limiting safety claims to technical performance~\citep{vertesi2015seeing}. This includes knowledge generated through technical upkeep and maintenance activities, which are often structurally underattended despite their proximity to conventional engineering concerns~\citep{russell2018after}. Such work frequently produces weak signals about accumulating risk that organizations oriented toward new features and technical performance may otherwise neglect. A safety culture can make it more legitimate for individuals to communicate these risks, question prevailing assumptions, and refuse to proceed when they judge failure to be foreseeable. Whether such actions alter system behavior, however, depends on the authority, incentives, and decision-making structures surrounding them. Yet being permitted to raise a concern is not the same as avoiding the outcome. The difference is whether the organization has agreed beforehand to treat this concern as decisive, rather than relitigating the issue each time.

\subsection{Curmudgeons \& Critics}
One practice in high-assurance engineering is to involve curmudgeons, critics, and ``pathological thinkers'' who attend closely to worst-case scenarios during development. This practice counters the ``optimism bias'' pervasive in Silicon Valley and other technology development settings~\citep{broussard2018artificial}. The point is not simply to introduce a negative perspective, but to assign people responsibility for questioning safety claims, identifying unsupported assumptions, and tracing plausible paths to failure.

Assurance cases provide one setting in which such roles can be made explicit~\citep{rushby2015interpretation,rismani2023beyond,bloomfield2024assurance}. An assurance case requires an organization to make bounded claims about a system's safety or fitness and to support those claims with specified forms of evidence. It can also assign responsibility for making, challenging, adjudicating, and revising those claims. Bloomfield and Rushby emphasize the role of skepticism throughout this process as a means of countering complacency and confirmation bias in AI system development~\citep{bloomfield2025ai}. One study found that a sociologist was particularly valuable in such a critical role~\citep{jatho2023concrete}.

The value of critics does not follow merely from assigning someone the role. The position of an ``official dissident'' creates a paradox if the organization treats criticism as another component to be added to a development process. Experimental studies of team dynamics often assign a participant the role of devil's advocate, but find limited effects when the dissent is not authentic or when other participants understand it as a procedural exercise~\citep{nemeth2001devil}. A critic who lacks independence, access to relevant information, protection from retaliation, or influence over decisions may provide the appearance of scrutiny without changing the conditions that produce risk.

Vaughan's study of the Challenger launch decision is a limit case. Organizations can undergo a ``normalization of deviance,'' accepting degraded states as normal when they produce no immediate visible consequences~\citep{vaughan1996challenger}. A few engineers repeatedly warned that launching below the rockets' designed temperature envelope raised the risk of failure, but pressures at both the contractor and NASA kept those warnings from controlling the decision~\citep{mcdonald2009truth}. A culture that permits dissent is necessary but insufficient: dissent also needs protected channels, standing, and real capacity to affect decisions.

\subsection{Traceable Processes}
A classical requirement in engineered systems is \emph{traceability}: the property that a system's design and behaviors can be tracked back to requirements or decisions in development, management, or operations~\citep{kroll2021outlining}. It supports failure investigation --- determining how a system reached a given state, or why it entered a hazardous one when control should have intervened. Unlike the weaker \emph{transparency} often called for in responsible AI, traceability attaches goals to information about a system, linking record-keeping and disclosure to review and analysis at specific stages and handoffs.

Physical systems often maintain a \emph{design history file} recording safety-relevant decisions. Large construction projects run on checklists that connect subcontractors to the project's state and record site changes for evaluation against specifications and to communicate between nonoverlapping specialty concerns~\citep{gawande2009checklist, ward2018effective}. Traceability attaches during operation too: aircraft carry flight data recorders (the ``black boxes'' investigators seek after accidents), ships carry voyage data recorders, and cars carry event data recorders.

Software, by contrast, is built for rapidly changing requirements through iterative lifecycles. Its tools produce reproducible version histories automatically, yet decisions about shifting requirements or goals are recorded, at best, in \textit{ad hoc} documentation --- even though software is more amenable to complete traceability than many engineered artifacts~\citep{kroll2021outlining, cobbe2021reviewable}. Only where software is highly regulated, as in medical devices or flight control, are requirements carefully traced to implementation.

An oft-suggested intervention to improve traceability is to provide documentation of decision-making for various portions of AI systems~\citep{heger2022understanding, mitchell2019model, gebru2021datasheets}. Such documentation is meant to increase risk communication, which our study of unlearned lessons from accidents suggests is valuable, but not a guarantee of safety. Documentation in service of traceability and failure investigation is a feature and necessary component of functioning safety processes, not a sufficient condition or a talisman that indicates safety processes are functioning properly. As artifacts, documentation objects can do very little on their own; yet they can be crucial if and only if they are part of an organizational practice through which people work through requirements, design choices, evaluations, deployment decisions, monitoring, incident response, postmortems, and future planning~\citep{geiger2018docs}. 

However, proposals for documenting datasets, model construction, and other facets of AI systems are often treated more instrumentally as component-level `solutions', often deployed outside of any substantive organization-wide governance process.  The widespread assumption that such tools improve risk communication and mitigation is, in fact, an empirical question requiring validation and study in the context of use. Documentation artifacts themselves (and the process of their preparation) can support effective risk perception, communication, and governance, but they can equally be ``expensive but meaningless rituals''~\citep{power1999audit}. Meyer and Rowan argue that making organizational structures and ceremonies more rigid can drive gaps between practices and policies or other stated goals~\citep{meyer1977institutionalized}. In the context of public accountability, Geiger et al.\ argue that documentation, transparency mechanisms, and targeted audits gain their power by exporting information to existing institutions, facilitating transparency if and only if they offer a method for incorporating surrounding stakeholder concerns into development concerns~\citep{geiger2024making}. As such, for traceability to function as a reliable safety practice, it must be integrated within a broader organizational safety culture, in which such documentation is part of the fabric of the organization and a medium through which teams work out issues, rather than something done retrospectively to check a box. 

These traceability and transparency practices can also be gamed, weaponized, and undermined through malicious compliance or symbolic compliance --- phenomena seen in domains from corporate law to education management to monetary policy to systems safety. A mountain of irrelevant and incorrect documentation can be worse than none at all. It is often the case that workers in systems are faced with a wide variety of rules, which might conflict or be subject to some interpretation about when they are meant to be applied. Thus, control at a system level can (paradoxically) be obscured by disclosure, since what seems like valuable communication can instead hide discretion or the accretion of risk when that communication is not clearly and systemically understood~\citep{strathern1997improving}. Alternatively, disclosure can be deployed purposefully for such an obscuring purpose, leveraging the talismanic effects of documentation to avoid real change, creating the symbols of compliance without decreasing (or even necessarily communicating formally about) risk, such as when organizations provide the bare minimum documentation needed to satisfy requirements for creating structures to limit sexual harrassment~\citep{edelman1992legal, edelman1999endogeneity}. This happens for benign reasons as well, such as trying to avoid perceptions of under-performance or risk over-acceptance. But it can also happen because of incentives, as we note above, which has been described as ``the tyranny of the bottom line'' or ``why good people do bad things''~\citep{estes1996tyranny}. Sampson argues that ``the `light' of transparency tools always creates `shadows' elsewhere'', in the context of understanding how and whether transparency mandates reduce corruption in international development programs~\citep{sampson2021good}.

\subsection{Inclusive, Diverse, and Psychologically Safe Spaces}
A safety culture must be inclusive and psychologically safe in ordinary practice, from routine operation to crises. The many parties in an AI organization --- workers, managers, contractors, domain experts, users, and impacted non-users --- bring heterogeneity that can breed interpersonal friction inhibiting good safety practice. \emph{Psychological safety}~\citep{edmondson_psychological_1999, edmondson_fearless_2018} is central: people participate, share information, raise concerns, and admit mistakes without fear of retaliation, discrimination, or social sanction. This literature more often measures productivity and profit than safety \citep{hong_groups_2004,herring_does_2009}, and while this link between diversity and safety has been theorized~\citep{foldy_power_2017}, it needs more empirical study. But safety decisions run on communication, and factors that inhibit it recur as antecedents to disaster (Section~\ref{sec:unlearned}). Interpersonal norms at the micro level, team and organizational structures at the meso level, and social institutions at the macro level can each either amplify these patterns or make space for the difficult work of safety.

Where psychological safety governs whether identified hazards can be raised, diversity governs which hazards are seen at all. Diversity matters not as a component dropped in to solve safety, nor a threshold past which disasters fall, but because people with different backgrounds, positionalities, expertise, roles, and tenure notice different hazards and feel compelled to raise different concerns, making different failure paths legible. Management science has long shown that diverse groups often outperform less diverse ones, especially on broad, open-ended, complex tasks.

Most pulse oximeters, for example, are far less accurate on darker skin, yet earlier studies deemed the error acceptable. Only in 2021 was this surfaced in a formal FDA alert, with much of the work done by Black engineers --- who comprise 5\% of U.S. engineers~\citep{mcfarling_poster_2022}. The AI parallels are in \emph{Gender Shades}, which found three major facial-recognition classifiers performed far worse on women of color~\citep{buolamwini_gender_2018}, and in the infamous case where Google Photos tagged pictures of Black people as gorillas \citep{noble2018algorithms}. We do not know the composition of the teams that created these products, but the failure to test on representative samples of humanity suggests the hazard --- and the U.S. tech industry is strikingly unrepresentative of the country, let alone the world~\citep{benjamin_race_2019}.

\subsection{Fostering Meaningful External Participation}
Beyond team diversity lies meaningful participation of stakeholders and impacted groups outside the organization, which is increasingly urged in AI safety as a way to surface concerns~\citep{Venkatachalapathy2025Participatory,lee2019webuildai,zhu2018value}. Wikipedia's ORES project, based on Participatory Design (PD) practices, let community members raise issues, audit beta models, tune parameters, and relabel training data --- resolving many biases and errors, but only because the whole surrounding socio-technical system was oriented to those ends~\citep{halfaker2020ores}.

Within Human-Computer Interaction, there is debate over whether ``Participatory Design'' (PD) practices can be a meaningful way of hearing from otherwise-absent voices, or a mere performance of ``pseudo-participation'' \citep{palacin_design_2020}, making designers feel empathetic and/or users feel heard, but without meaningful change. Much of the PD literature holds that such practices are not a panacea for participation, as meaningful involvement requires redistribution of power, not merely consultation \citep{arnstein_ladder_1969, birhane_power_2022}. An empirical study of ostensibly participatory stakeholder involvement in the U.S. AI industry found such participation is dominated by extracting commercially-relevant concerns, overwhelmingly begins late in the development process, and rarely incorporates impacted non-users \citep{kallina_stakeholder_2025}. Like documentation, audits, and benchmarks, participation can surface safety-relevant information --- but only if the surrounding socio-technical system is configured to support it.

\subsection{Failure is Normal}
Safety work cannot depend on the hope that harmful states will be rare or self-evident. Disasters often ``incubate'' \citep{turner1978man} over long periods. High-reliability organizations counter this incubation with a ``preoccupation with failure'', treating anomalies and near misses as information about the state of the system rather than as noise \citep{weick2015managing}. Organizations responsible for AI systems should likewise expect failures and near misses, and rehearse how they would recognize, communicate, mitigate, and learn from them.

Simulations, incident exercises, and pre-deployment disaster scenarios surface latent assumptions while there is still time to redesign or strengthen controls. In a premortem, a team assumes a project has already failed and generates reasons, legitimating reservations members would otherwise not voice~\citep{klein2007premortem}. In ``chaos engineering,'' operators inject failures into production to test resilience and rehearse recovery~\citep{basiri2016chaos}. As ever, these work only inside a safety culture; deployed as component-level `solutions,' they become hazards themselves if people take them to have `solved' safety. These lessons generalize across technology stacks: the tech industry's own ``Site Reliability Engineering'' discipline institutionalizes many of them, which we take up in Section~\ref{sec:agenda}, where SRE anchors our argument that AI is not exceptional~\citep{beyer2016site}.

\subsection{Structural Incentives for Meaningful Self-Regulation}
\label{sec:selfregulation}

A longstanding and controversial theory holds that industries generating societal risk have an incentive to self-regulate, building legitimacy while keeping autonomy from regulators. Supporters call this ``responsive regulation''~\citep{ayres_responsive_1992}, resting on an implicit ``Social License to Operate'' publics grant or withhold~\citep{slo2016}. Critics counter that self-regulation is often a tool to delay or weaken oversight. Lacking enforcement, it slides into superficial compliance or regulatory capture, and into ``symbolic compliance''~\citep{christmann2006firm, edelman2011comply}, where firms adopt the appearance of safety through voluntary codes and CSR without changing core practices or business models.

This final element of safety culture is largely external to the organization. Even well-designed practices are overwhelmed when the political economy rewards speed, scale, opacity, or symbolic compliance over real risk reduction. Meaningful self-regulation therefore depends on structural incentives that make safety action consequential, protect those who raise concerns, and impose costs for ignoring documented hazards. Other sociotechnical industries have faced this. Nuclear power, chemicals, aviation~\citep{oster2013analyzing}, and finance were transformed by recognizing that their survival was bound to the safety of their practices~\citep{gunningham1997industry}. Omarova argues that only by enlisting finance's active participation in regulation can meaningful reform escape the loophole-closing cat-and-mouse game~\citep{omarova2010wall}. That recognition is, at best, uneven in the tech industry and AI research. Where a firm has a longer-term business interest in binding itself, such as avoiding downtime and data loss, the commitment can be internal. Yet where safety cuts against short- and long-term interests, the binding must come from external forces like regulation.

Whether self-regulation is meaningful or symbolic hinges on the freedom of experts to speak and act inside their organizations, often against dominant incentives. Safety culture legitimizes such action, but the capacity for it is shaped by the individual's relationship to employer and field. Because the market for AI developers is fiercely competitive, leading experts can choose firms where they feel best able to meet their goals, functional and safety alike, and do move on perceptions that a firm is serious --- or unserious --- about safety.

Worker mobility raises questions of intellectual property, including trade secrets about safety practices. One might expect IP protections and contracts to bind safety-minded workers to their employers' economic incentives; Keller and Aplin argue instead that democratic governance should shape the scope of IP in US and EU law, opening space for targeted disclosures about how AI systems behave~\citep{keller2026reconciling}. This foreshadows protected-reporting regimes in aviation~\citep{chidester2007voluntary} and medicine~\citep{pope2012physicians, boysen2013just, shimabukuro2015safety}: structural incentives that make surfacing hazards safe for the individual and consequential for the organization. Without them, the safety-culture elements above rest on voluntary commitments that the incentive problems of Section~\ref{sec:unlearned} routinely undermine.

\section{Making Safety Sociotechnical: A Research Agenda for Managing AI Harms}
\label{sec:agenda}
A safer AI
involves more than technical interventions to components (models, datasets, assessment practices). AI safety demands adopting processes, reflexive governance systems, and organizational structures that are responsible and accountable to the interests of the most vulnerable. It requires social understanding of how large-scale AI systems interact with existing unjust social structures. There are several nascent lines of research that begin this work~\citep{raji2020concrete, raji2021ai, raji2022fallacy, dobbe2022system, raji2022algorithmic, rismani2023beyond, shelby2023sociotechnical, jatho2023concrete, nouws2023diagnosing,sloane2022silicon, geiger2024making}. And we stress again that this social understanding not only implicates technical fixes to known risks, but also establishing maintenance programs where human oversight is not individualized but part of a governance system~\citep{vinsel2020innovation}. This is the philosophy behind systems-theoretic approaches to safety, moving beyond fault tolerance and reliability-driven approaches.
\looseness=-1

Recent research~\citep{rakova2023algorithms}, nascent regulatory frameworks~\citep{nistairmf, golpayegani2023high, cobbe2023understanding}, and industry initiatives~\citep{gebru2021datasheets, mitchell2019model} are attempting to grapple with the potential safety and social risks of the widespread use of AI, often seeking new component-level interventions like checklists, benchmarks, impact assessment processes, safety models, and more. These efforts, which are often combined into ``responsible'' or ``trustworthy'' AI, are critical. However, the very existence of such terms reveals a structural absence: we do not discuss ``responsible civil engineering'' or ``trustworthy medicine'' in the same manner. Those fields also have countless component-level technical safety interventions, but also well-established internal and external governance mechanisms and cultures that generally make such safety interventions not just enforceable, but normative.

\subsection{Site Reliability Engineering as a Sociotechnical Precedent}
As an example of how these techniques can defeat the argument that AI or software systems are somehow different to other engineered systems, we offer the example that many large technology platforms achieve infrastructure-level reliability despite being complex, ever-changing, and prone to emergent failures~\citep{garraghan2018emergent}. This high performance is the result of a sociotechnical and organizational outlook toward risk, combining technologies that make operations safe (i.e., non-failing) with a safety culture that values measuring and managing risk by aligning it to business goals. The industry refers to this discipline as ``Site Reliability Engineering'' (SRE)~\citep{beyer2016site}. As in other safety-intensive domains, SRE deploys engineering methods to avoid defined losses (e.g., unavailability of an internet service, data loss, unacceptable variability in latency or throughput) and combines these with organizational actions to identify and communicate risk information and pass it to decision-makers for action. Part of the work of these staff is to inculcate a safety culture among engineers working on improving system function as well as among managers and executives, shaping their organization's outlook towards controlling risks (e.g., by shaping and clearly defining accountabilities for risk tolerance, such as setting thresholds on when new product features can be launched to avoid disruption, and creating a culture where delaying launches when these thresholds are violated becomes acceptable despite the associated costs). Reporting on an informal study of staff at Google who had worked in safety-critical industries, some creators of this doctrine note four main themes for high-reliability operation across disciplines:
\begin{inparaenum}
\item Preparadness and Disaster Testing;
\item a ``Postmortem Culture'';
\item Automation to reduce operational overhead; and
\item Structured and Rational Decision-Making~\citep[Ch. 33]{beyer2016site}.
\end{inparaenum}
These align to our unlearned lessons about poor risk perception, avoidance of negative information (``no bad news''), and permanent rush cultures.

SRE differs from safety in being a discipline of risk management, explicitly embracing ``the right amount'' of failure and focusing on high-reliability goals, vs. safety's focus on avoiding defined losses entirely. The \emph{error budget} around uptime, SRE's central instrument, works not so much because high availability is naturally aligned with all interests, but more because SRE has been a political movement to get leadership to commit in advance to a more long-term valuation of a reputation of reliability, versus more short-term pressures to ship products. While the canonical SRE textbook argues the error budget ``removes the politics from negotiations between the SREs and the product developers'' \citep{beyer2016site}, it rather relocates and formalizes it. Leadership often retains an override, but this is deliberately expensive within SRE culture and requires postmortems. We offer this example of how adopting a sociotechnical response to risk can tame incentives like those to always ship and gives a generalized resource for engineers to push back---but does not eliminate risk and is contingent on leadership committing to this priority. Where such interests are not aligned, exogenous factors like regulation and stakeholder pressure might nudge leadership towards such committments.  

\subsection{Who Governs Safety?}
The field of AI must broaden its horizon of available solutions beyond technical component-level practices to sociotechnical system-level interventions, as the efficacy of such practices on real-world harms is a system-level property contingent on higher-order governance structures. What use is a voluntary bias benchmark or red team security audit if leadership routinely decides to ship products even if those tests fail --- especially if there are miles of passing results from other technical tests across the responsible AI literature? Benchmark results and audit findings may have some utility, such as evidence for a public whistleblower, but then the efficacy of such tests for safety depends on the efficacy of whistleblowing in a given institutional context. 

This argument is about jurisdiction, as we critique the status quo in AI research, development, and media coverage that turns to the unrepresentative set of elite technical developers and experts for answers to societal concerns with the products derived from their inventions \citep{benjamin_race_2019}. There is a long tradition in which technocrats and titans of industry like Henry Ford and Robert Moses have used their technical authority to assert control over the social implications of their inventions \citep{winner_whale_1986}. First, such jurisdictional claims reinforce the existing capture of all kinds of social, political, and economic institutions by technologists and the tech industry \citep{pasquale2015black, cohen2019between}; 
second, the origins, impacts, and mitigation strategies of the harms arising from high consequence systems of all kinds---including AI systems---are all fundamentally sociotechnical, so purely technical expertise cannot replace sociotechnical expertise in mitigating harms arising from AI \citep{friedman2019power}.
To ignore or marginalize experts in the field of sociotechnical systems and systems safety in favor of technocratic experts is itself a recipe for disaster. However, they persist in part because technical components that promise more narrowly-framed safety solutions can be more easily evaluated with quantitative metrics, such as an Area Under the Curve (AUC) metric for measuring, for instance, how well a model trades off harmful vs. useful output from a given set of inputs. 

Practical safety practices around AI systems requires learning the lessons identified in \S\ref{sec:unlearned} and making those learnings live in organizations which control those systems. In a systems safety lens, properties such as model performance, the expression of bias, robustness, conformance to human values/alignment, maintainability, etc., can be operationalized at the component level and at the system level. Much existing work aims to ``embed'' these properties in the design of technical artifacts. However, in many high-assurance domains, safety is part of an entire sociotechnical system and its assessment requires contextualizing requirements on component processes Achieving this requires both basic and applied research, for which we offer a high-level agenda here. 

\subsection{Operationalizing Risk Frameworks}
It is critical to engage in this research now, as the structures for risk perception and management and the safety systems which will manage harms arising from AI systems for the next epoch of technology governance are forming and ossifying now, via the creation of policies and practices. Consider emerging tools for organizing this work across organizations, such as US NIST AI Risk Management Framework.\footnote{While the NIST framework is offered as a voluntary tool and not a standard, many standards bodies---the International Organizations for Standardization (ISO), the Institute for Electical and Electronics Engineers (IEEE), and the International Telecommunications Union (ITU) are all independently pursuing standards-track approaches meant to create standard controls on risk in AI components and systems. 

While the NIST framework offers guidance at the organizational level, the in-process standards are much more instrumental and component-focused. Standards such as ISO 420001 describe system-level management, but how to ensure they function effectively in practice remains an open research question.} Although many organizations have their own frameworks for thinking through AI risk concerns, the NIST framework has emerged rapidly as the focal point for industry, government, and public policymaking. However, while the framework attempts to define risk management as a holistic organizational activity, it explicitly espouses flexibility and eschews a prescriptive approach. To make use of it in any specific application or organization, therefore, requires substantial interpretation and specialization. As noted among our unlearned lessons, poor risk perception is a recurring cause of undesired outcomes, and risk discovery activities are at risk of becoming empty rituals which promulgate the sense that risk is managed decoupled from actual risk information or control. Emerging standards apply largely to behavioral aspects or reliability of specific components, without covering system-level interactions or defining system-level outcomes. Safety governance systems in many high-assurance domains have clear metrics that can justify judgments about assurance of technology use for particular purposes or of programmatic success in safety control. How can we develop similar empirical grounding for safety processes around AI systems?
\looseness=-1

It is straightforward to see the open questions around applying the NIST AI RMF as demanding new checklists, system metrics, and other technology-focused activities and structures responsive to the concerns in the framework. And while such activities are critical, even more important is operationalizing the concerns the framework articulates into an entire risk management system capable of identifying system-level problems arising both within the AI system under management and between that system and its environment. At present, few if any risk management  approaches in computing are known to have this property. For example, checklists are best understood as tools for risk elicitation and risk information communication, but in computing are often treated as dispositive performance standards; code in other clothing. Yet if there are no consequences to failing to adhere to a checklist, what is its utility? Research is needed to understand how assurance can be reached for technical components and then joined to risk assessment at the system level informing governance programs as situated within controlling organizations. Policy must drive this posture and create enforcement mechanisms capable of determining when risk perception does not empower risk mitigation. NIST itself aims to open these research questions to validation through a testbed aimed at viewing risks through a sociotechnical lens~\citep{nist-aria-2024}.

Validation of new governance practices is a critical open research-level problem. Consider the heavy focus of recent research on structuring the documentation of AI system capabilities, limitations, and component provenance (e.g., dataset documentation)~\citep{gebru2021datasheets, mitchell2019model, holland2020dataset, winecoff2024improving}: despite many competing standards, there is little research examining the extent to which such practices improve risk information validity \& communication for the purpose of improving system safety either within the organizations controlling or developing the AI systems or between those organizations and broader sets of stakeholders such as regulators and publics. Linking concrete governance practices to meaningful safety improvements (and eventually further up the hierarchy of system control to the level of organizational and public policies~\citep{chen2024explainer}) remains a critical thrust for research on improving the safety of AI in sociotechnical systems.
\looseness=-1

Our call to reframe AI safety as a sociotechnical problem requiring consideration of full systems rather than component properties echoes other recent work. This work is unique in its focus on learning from safety failures across domains to extract core structures by which the cognizant organizations responsible for the system can achieve safety. But other work aims to situate technical decision-making and questions of operationalizing values in context~\citep{kudina2024sociotechnical} or to define governance characteristics by recognizing the ways in which AI systems can exhibit the properties of complex, adaptive systems~\citep{kolt2025lessons}. Others have made more specific interventions by exploring the sociotechnical nature of risk in specific domains, such as aviation~\citep{hayes2025rigid}, medicine~\citep{mccradden2023normative}, nuclear power~\citep{verma2024sociotechnical}, government services~\citep{andrus2020ai, dobbe2024toward}, and military applications~\citep{jatho2022artificial}. 

\subsection{Applying Systems-Safety Methods to AI}
However, an interesting open direction is to articulate the insufficiency of component-level analysis methods on their own (e.g., alignment, interpretability, uncertainty quantification, robustness enhancement, causal reasoning, fairness measures) and to recompose that critical decomposition into a positive framework for addressing risk and safety assessment across entire sociotechnical systems. Some existing work takes on portions of this effort: scholars have investigated the limitations of instrumental explanation~\citep{miller2019explanation, doran2017does} and described how to view bias and fairness definitions with robust context~\citep{mulligan2019thing, blodgett2020language, mitchell2019model}. Khlaaf argues strongly in favor of viewing methods for requirements refinement, such as alignment or fine-tuning, as descriptions of functional behavior rather than safety interventions~\citep{khlaaf2023toward}. However,  there remains much leverage in validating the risk reduction potential of auditing and other evaluation methods, including techniques such as red teaming, connecting component-level interventions to empirically-grounded safety claims. Our project in this work is to ground the need for and to scope the goals for such efforts.

One promising research thread in that direction is to apply existing methods from systems safety to the problem of AI system risk perception and reduction~\citep{dobbe2022system, rismani2023plane, khlaaf2023toward}. For example, emerging research applies explicit systems-theoretic modeling of accidents to understand the ways AI systems can fail in their social and ethical context~\citep{raji2020concrete, raji2022algorithmic, jatho2023concrete, dobbe2024toward, delfos2024integral}. This research approach explicitly disclaims the common use of the term ``AI Safety'' to refocus questions of harm avoidance in systems where AI is involved toward practices and structures that manage risks borne of those systems in the real world.

The use of Leveson's System-Theoretic Accident Model and Processes  (STAMP) model, for example, has been called out explicitly as a useful tool which drives in this direction~\citep{dobbe2022system, jatho2023concrete, rismani2023beyond, rismani2023plane, delfos2024integral}. STAMP aims to increase safety by engineering around defined failures ("system losses") at the system level, modeling systems as hierarchies of sociotechnical control and explicitly scoping regulatory, organizational context, and management into the engineering analysis~\citep{leveson2016engineering, leveson2023introduction}. One motivation for developing STAMP was to move beyond probabilistic risk assessment due to the difficulty of accurately modeling probabilities. As Leveson argues, probability is unknowable. Specifically, STAMP relies on identifying where components of a system can exert "control actions" on other components and receive feedback. Building a graphical model of a system, enriched where necessary by quantified control dynamics, enables engineers and safety analysts to search a specific structure for paths to hazards (being states of the system that immediately prefigure losses in worst-case environmental conditions). With hazards identified, it becomes easier to determine interventions such as offsetting control requirements (to assure that a system will exit a hazard state safety and recover normal operations) or redesign to eliminate the hazard altogether (STAMP describes such redesign as the gold standard for safety intervention). 

By examining safety at the system level, STAMP aims to model component interactions and resulting emergent hazards explicitly. Yet Perrow and the high-reliability literature agree that accidents are emergent from complexity, but disagree on whether added controls are helpful or just add more complexity and thus more risk. Leveson's response, which we follow, is that added control can be beneficial when it is fully derived from and checked against an explicit model of the system, rather than bolted on and drifting constantly out of sync with the real system---hazard-eliminating redesign is the limit case that lowers risk and complexity at once~\citep{leveson2009moving,perrow1984normal, roberts1990managing}.

An advantage of STAMP is that it has been used across a variety of similar sociotechnical domains to identify hazards proactively and recommend system-level interventions, controls, and redesigns to support the avoidance of bad outcomes~\citep{leveson2004role, leveson2011applying, leveson2012subsafe}. Important open questions include: can the utility of STAMP translate to AI systems? Do the tools and techniques of this method require changes to deal with the problem that AI systems are often specified inductively rather than procedurally? Does an encoding of social and ethical risks and harms into the language and tools of safety engineering limit the rich, thick view of these problems desired by activists, researchers, and policymakers focused on human values? All models are wrong, but some are useful, and STAMP is a model that will always only partially represent a system. An analysis is only valid inside the frame it draws, substituting proxies for constructs that matter. The more detailed this interior frame is, the greater the risk that passing a STAMP analysis is taken for a safe system. So how can we productively mind this gap?  We consider research addressing these questions to be a critical next step in the realization of sociotechnical management of the responsible use of AI, and a key enabler of computing technology advancement which benefits society.
\looseness=-1

Many organizations have invented or adapted frameworks to attempt this type of work, so an important research line is to study the performance of those methods \emph{in situ}. All studies of risk reduction suffer from a critical epistemic blindness: one tail of error cannot be observed (i.e., problems experienced conclusively demonstrate failure, but non-experiencing of problems does not guarantee their absence). Research along these lines suggests that practitioners struggle to close the gap between instrumental, component-level controls and system-level hazards~\citep{rismani2023plane}, may prefer to intervene at locations in the system not considered by available techniques~\citep{holstein2019improving}, struggle to make use of internal tools meant to operationalize abstract concerns such as safety and ethics~\citep{madaio2020co, morley2023operationalising}, and report that organizational structures and incentives hinder their work~\citep{rakova2021responsible}. Broadly, these empirical studies validate our unlearned lessons as the key barriers to improving safety in AI systems and ground our research agenda firmly in the tools, techniques, procedures, and system structures required to overcome them reliably, as happens across high-assurance domains.

\section{Conclusion: History Doesn't Repeat Itself, But It Rhymes}
Ralph Nader's famous 1965 book \emph{Unsafe at Any Speed} was a clarion call to reframe the concept of ``safety'' in the automotive industry to capture a focus on reducing harms to vehicle occupants and surrounding pedestrians. The then-standard mantra from the auto industry --- the ``three E's: Engineering, Education, and Enforcement'' --- framed safety problems away from vehicles to avoid taking responsibility for the ways system design could cause harm. (Education and Enforcement referred to modalities of improving driver behavior, while Engineering was intended to cover the design of roads rather than vehicles: that is, the industry shaped the reference frame of ``safety'' so as to include only things not under the control of car designers, manufacturers, and sellers.)
Critically, Nader’s book redefined safety to consider vehicles as used in their environments, consistent with the modern safety engineering view that safety can only be evaluated in the context of use. Because safety arises from the affordances and capabilities of their systems/artifacts, this broader notion of safety included, e.g., that vehicle safety analysis must not only consider the protection of the vehicle in a collision, but the ``second collision'' of the occupants within the vehicle itself---a harm well-mitigated by now-standard and standardized seatbelts.

We offer that a version of Nader's critique applies to the AI industry today: ``AI Safety'' 
is, and ought to focus on, the real harms playing out now that disproportionately impact already-marginalized communities, in the context of the actual complex socio-technical systems in which technical components operate.
In AI, the focus is often on improving performance even when this can add risk, such as when improving object detection and classification or reducing model bias improves surveillance capabilities, increasing the opportunities for harm.
It is easy as well to confuse performance improvements with safety improvements - classifying objects more reliably, for example, means fewer misclassifications; if LLMs generate more factual information, the chance to mislead a user is lower.
But reliability is not safety, and at any level of performance or for any tradeoff between failure modes (``at any AUC’’) a system that causes harm is unsafe.
Even when improved performance (e.g., fewer misclassifications) does appear to increase component reliability, there is no guarantee this corresponds to safer behavior at the system level (e.g., when more accurate classifications perfect automated surveillance).

Interventions are currently more offered at the component level and focused narrowly on industry-aligned goals such as robustness (analogous to improving the collision survivability of vehicles at the cost of collision survivability of passengers) and improved performance (seatbelts and other safety features can just as well lead to \emph{riskier} behavior by operators, who use safety technology to equalize their risk tolerance)~\citep{latour1992missing}. Nader also argued for stronger government intervention in vehicle safety. Regulation can be a productive component of reducing AI harms, but smart regulation must demand actions that improve AI controllers' and deployers' capacity to perceive and act on the risk of harms. Careful reflection on the lessons of past disasters combined with a sociotechnical system-level view of AI technologies is critical to shaping safer outcomes.
\looseness=-1

Safety, the engineering effort devoted to ensuring a system does not exhibit defined undesired behaviors, must not be confused with related conceptual framings (``responsible use'', ``risk management'', etc), which have substantially different political economy profiles. The acceptability of an artifact for use in a particular application depends on the definitions of losses (Nader's book argues strongly in favor of losses previously dismissed by the automotive industry and regulators, for example harms to pedestrians outside the vehicle and the ``second collision'' of vehicle occupants with the vehicle itself, which is what causes injury). Risk management, by contrast, is about tradeoffs, setting thresholds of the form ``As Low as Reasonably Practicable''~\citep{MELCHERS2001alarp} and may serve to launder acceptability through visions of control even when control is reduced by financial or other equities that raise tolerance for risk. Safety is not neutral: opening the definitions to democratic discussion is a precondition to identifying the tradeoffs necessary for governance~\citep{abdu2025everyday, cohen2019between}.

Following Nader's efforts to raise choices within safety to the level of visible discussion around fitness-for-use of new technologies, we make a similar argument here, pushing beyond existing calls for responsible AI~\citep{Leslie2020Tackling, Banks2024Data}: AI safety must consider not only properties of models but the safety of models as components in larger sociotechnical systems (for which controls against failure might come from outside the models themselves). This approach shows substantial progress in improving the fitness-for-purpose of AI systems in high-consequence applications~\citep{dobbe2024toward, delfos2024integral, rismani2025measuring}. This is not merely a question of ``responsibly using'' technology, but a question about evaluating the consequences of using a tool in a use case for a purpose against a defined set of undesired outcomes. Translating these social and ethical questions to design requires careful analysis of full sociotechnical systems.

Safety claims are quantified differently: over \emph{all} foreseen system and environmental states, the system will avoid losses. (Of note: safety cannot be claimed in an absolute sense, only falsified; here we mean the claim is quantified universally with regard to the assumptions and system model relied on by safety analysis, distinct from risk management in that the goal is elimination of loss rather than bounding it below some tolerable probability.)

We call on industry and the research community to reimagine ``AI safety''. Safety is a property of complex socio-technical systems. It results from social structure and hierarchy, communication and interaction. Socio-technical systems cause their own behavior, and they contain not just technical and human components that can be ``fixed'' to be reliable or unbiased or compliant with a specification, but also the social, organizational, cultural, and economic context in which those components act in concert. System behavior safety exists only for this composed assemblage; without a context of use, claims about safe (responsible, reliable, trustworthy, unbiased, etc.) component performance are merely technical artifice --- a dial setting against the abyss. The boundary between a system and its environment is often fungible and resistant to firm definition. 

A systems view includes an understanding of policies and other normative constraints, and also the reaction of components (human and technical) to these. A systems view grounds and contextualizes the value of interdisciplinary research and interventions by elucidating the interfaces along which problems can arise. As with automobiles, the goal of safety cannot be to eliminate component failures or even crashes---instead, negative events must be survivable transitions of entire systems. The major lesson from analysis of disasters from a sociotechnical lens is that individuals (developers, operators, or idealized developer-operators) can make bad decisions, but that such decisions lead to accidents only when the system's structure allows~\citep{leveson2011applying}. The factors we mention in  \S\ref{sec:unlearned} have all contributed to man-made catastrophes, and we see these same hazards in AI research and industry. The future of responsible AI requires learning from past catastrophes.

\printbibliography

\end{document}